# Life Beyond the Solar System:
# Observation and Modeling of Exoplanet Environments

*A white paper submitted in response to the NAS call on the Astrobiology Science Strategy for the Search for Life in the Universe*

Lead author:

Anthony Del Genio
NASA/Goddard Institute for Space Studies, New York, NY
Phone: 212-678-5588
Email: anthony.d.delgenio@nasa.gov

Co-authors:

Vladimir Airapetian (NASA/GSFC and American University, DC)

Dániel Apai (University of Arizona, AZ)

Natalie Batalha (NASA Ames Research Center, Moffett Field, CA)

Dave Brain (LASP/University of Colorado)

William Danchi (NASA/GSFC)

Dawn Gelino (NASA Exoplanet Science Institute, Caltech)

Shawn Domagal-Goldman (NASA Goddard Space Flight Center, Greenbelt, MD)

Jonathan J. Fortney (University of California, Santa Cruz)

Wade Henning (NASA/GSFC and University of Maryland Dept. of Astronomy, College Park MD)

Andrew Rushby (NASA Ames Research Center, Moffett Field, CA)

The search for life on planets outside our solar system has largely been the province of the astrophysics community until recently. A major development since the NASA Astrobiology Strategy 2015 document (AS15) has been the integration of other NASA science disciplines (planetary science, heliophysics, Earth science) with ongoing exoplanet research in astrophysics. The NASA Nexus for Exoplanet System Science (NExSS) provides a forum for scientists to collaborate across disciplines to accelerate progress in the search for life elsewhere. Here we describe recent developments in these other disciplines, with a focus on exoplanet properties and environments, and the prospects for future progress that will be achieved by integrating emerging knowledge from astrophysics with insights from these fields.

This is one of 5 "Life Beyond the Solar System" white papers submitted by NExSS. The other papers are: (1) *Exoplanet Properties as Context for Planetary Habitability;* (2) *Technology Development for Continued Progress;* (3) *Remotely Detectable Biosignatures;* (4) *Exoplanetary Space Weather and Habitable Worlds.*

## 1. Areas of significant scientific progress since publication of the NASA Astrobiology Strategy 2015

*A. Exoplanet Observations*

Most recent observational advances are covered by the companion *Exoplanet Properties* white paper. Here we note major developments since AS15 that are relevant specifically to exoplanet environments and habitability:

i. The first identifications of potentially habitable nearby Earth-size planets: Proxima Centauri b [1], TRAPPIST-1 e (and perhaps f,g) [2], and LHS 1140 b [3].

ii. The discovery that atmospheres can be retained on short-period rocky exoplanets despite a hostile stellar environment (GJ 1132 b) [4].

iii. A gap in the radius distribution of close-in Kepler planets that documents planets with atmospheric photoevaporative loss and planets that formed gas-poor [5].

iv. Mass-radius diagrams for planets with both RV and transit detections that differentiate rocky, Neptunian, and gas giant planet regimes [6].

v. Categorization of planets by susceptibility to various escape mechanisms to infer exoplanets with vs. without atmospheres [7].

vi. Possible first evidence for ongoing geological activity/surface lava exposure on an exoplanet (55 Cnc e) [8].

*B. Exoplanet Modeling*

i. Since AS15 there have been advances in identifying false positives for biosignatures, frameworks for biosignature assessment, and statistical/quantitative approaches. The NExSS Biosignatures Workshop in 2016 led to 5 papers on these topics. See the companion *Remote Exoplanet Biosignatures* paper for more details.

    ii. The concept of habitable planets has been expanded to include greater consideration of high-energy radiation from the host star, in particular ionizing radiation and stellar wind particles and the magnetic field (space weather) and their impacts on atmospheric chemistry and escape [9], [10], [11].

    iii. At the time of AS15, most estimates of exoplanet habitability were based on 1D model studies. Since then, 3D Earth global climate models (GCMs) have become widely used tools for understanding factors determining climates and thus potential habitability of exoplanets. These include the role of oceans [12], [13], [14], effects of partial land coverage [15], [16], cloud stabilizing effects as a function of planet rotation [17], [18], and differences between planets orbiting cool vs. warm stars [19], [20]. 3D models are now routinely used to create synthetic transit transmission spectra and phase curves [16], [19], [21], [22] to inform the interpretation of exoplanet observations.

    iv. Since the New Horizons flyby of Pluto in 2015, considerable evidence from observations and modeling implies that long-duration high-volume subsurface water oceans are a very common feature for icy trans-Neptunian object analogs, icy outer moons, and planets not bound by host stars (Nomad planets) [23], [55], [56], [57].

## 2. Promising key research goals in the field of the search for signs of life in which progress is likely in the next 20 years

*A. Factors affecting the potential for life on planets orbiting M dwarfs*

    Most rocky planets that will be discovered by missions such as TESS and ground-based telescopes (e.g., MEarth, E-ELT), or characterized by transit transmission spectra acquired by JWST, in the next 20 years will be close-in tidally locked objects orbiting cool stars. The highest priority for cross-discipline studies will be to understand the physical processes that make these planets more or less promising for habitability than more Earth-like planets orbiting warmer stars, to determine optimal M star candidates that minimize the deleterious effects of the stellar environment, and to begin to characterize rocky planets orbiting such stars. Steps to achieve these goals include

    i. Modeling the impact of pre-main sequence elevated luminosity [24], [25] and energetic particles [9], [10] on the retention of atmospheres, and specifically water [26], for a variety of planets.

    ii. Understanding the factors that determine the strength of tidal heating and global magnetic fields and their impact on exoplanet habitability [27], [28].

    iii. Constraints on spin synchronization: Recent research utilizing advanced geophysical interior modeling finds capture into exact 1:1 spin-orbit resonance may not

be as common or inevitable as previously thought for close-in terrestrial planets [50], [54]. Equilibrium at higher order (e.g., 3:2) spin-orbit resonances must be given strong consideration in climate models [14]. Pseudo-synchronous states (e.g., 3% faster than synchronous) are now believed likely for worlds with liquid layers [52], and unlikely for solid worlds [53]. Most importantly, atmospheric torques even for thin atmospheres are now found to be sufficient to break typical 1:1 spin-orbit resonances [29].

    iv. Modeling the effects of ionizing radiation of planet-hosting stars on the initiation of prebiotic chemistry on rocky exoplanets [11].

*B. Greater utilization of short time scale 3D global climate models and geologic time scale carbonate-silicate cycle and planetary interior models to simulate the potential habitability of exoplanets for all stellar types*

    i. 3D GCMs will be applied in several ways: (1) Broadly, to understand the limits of and optimal conditions for habitability [18], and to help prioritize promising targets for characterization among the large number of habitable zone planets found by TESS and ground-based telescopes. (2) Narrowly, to simulate the detectability of biosignatures for specific exoplanets of interest (TRAPPIST-1, Proxima Centauri b, LHS 1140 b etc.) [16], [21], [22]. (3) To explore potential synergies between the more easily characterized atmospheres of giant exoplanets [30] and those of harder to observe rocky planets. The goal will be to identify features that are robust across a variety of models.

    ii. Planetary evolution over geological/astronomical time will be quantified, including changes in radiogenic heat flux [31], modes of tectonism [32], tidal activity [27], [28], and associated effects on long-term planetary habitability [33], [34]. This includes improving understanding of interior-atmosphere connections [35], tidal-orbital feedback [27], and how planets with greater or less geologic activity (e.g., volcanism), or differing modes of tectonism (e.g., plate tectonics vs. stagnant-lid) lead to variations in habitability by altering chemistry, surface temperature [34] or surface water availability [36]. Questions about how the carbonate-silicate cycle operates on geologic time scales to regulate surface temperature [37], whether it can efficiently regulate $CO_2$ on global ocean "aquaplanets" [38], and whether aquaplanets have reduced primary productivity [39] that affects biosignature detectability [40] will need to be addressed.

    iii. Solar system objects will be used to test and constrain exoplanet models and theories. Examples include global magnetosphere-ionosphere-thermosphere models to study the impact of extreme space weather on Earth, Mars and Venus [11], [41], the "cosmic shoreline" theory of atmospheric retention [7], tidal maintenance of subsurface water on Europa and Enceladus [42], [60], and mechanisms that may explain habitable conditions on the terrestrial planets under the faint young Sun [43], [44], [45], [46], [47].

iv. Laboratory work will be conducted to better understand atmospheric processes and their effect on temperature structure and spectra, especially for extreme temperatures and pressures and exotic species characteristic of hot gaseous planets and the early histories of rocky planets:  Gas opacities, chemistry, cloud formation [48].

v. Laboratory experiments to determine whether energetic particles and cosmic rays can compete with known mechanisms for the formation of hazes and clouds [49].

*C. Biosignature studies*  (see the companion *Remotely Detectable Biosignatures* paper).

**3. Key technological challenges in astrobiology as they pertain to the search for life on exoplanets**

i. It has now become feasible to conduct large ensembles of 3-D Earth climate model simulations. Such ensembles for exoplanets, sampling the full range of potentially observable external parameters (rotation, instellation, etc.) as well as difficult to observe internal parameters (surface pressure, greenhouse gas concentrations, etc.) may help to identify the parts of the space most favorable to habitability, perhaps with the aid of machine learning approaches that detect optimal parameter combinations.

**4. Key scientific questions in astrobiology about the search for life on exoplanets**

i. What biogeochemical and climatic factors determine the robustness and remote detectability of exoplanet life? A wide range of environments may support life, but to detect it remotely, we need to understand the processes that favor abundant life with strong, unambiguous biospheres.  These include interior processes that control geochemical fluxes of gases, biosphere robustness on aquaplanets vs. planets with continents, external parameters that control the spatial extent of clement temperatures and surface water availability, conditions that produce clouds or hazes that will confound attempts to detect biosignatures, and false positives that may mask biotic signals.

ii. Do planets build up thick $CO_2$ atmospheres through the carbonate- silicate cycle feedback at cold temperatures? The outer edge of the habitable zone for "Earth-like" planets is defined by extrapolating the carbonate-silicate cycle feedback that seems to occur on Earth to cold temperatures [58], implying the buildup of many bars of $CO_2$. A challenge for the coming two decades will be to observationally constrain the extent to which this process operates on planets irradiated more weakly than Earth [51].

iii. What atmospheric properties (e.g., temperature structure, chemical abundances, circulation regime) of more easily observed large exoplanets (Jupiters to sub-Neptunes) can be observationally constrained by near-future missions? These planets may in some respects serve as a testbed for theories about the atmospheres of rocky exoplanets that are more challenging to observe [14], [59].

**5. Scientific advances that can be addressed by U.S. and international space missions and relevant ground-based activities**

 i. TESS will greatly expand the population of known potentially habitable exoplanets, some of which may be selected for atmospheric characterization by JWST mid-IR transit transmission and eclipse spectra to search for $H_2O$ or biosignature gases. Other JWST contributions are discussed in the *Exoplanet Properties* white paper.
 iii. Extreme Precision (10 cm s$^{-1}$) Radial Velocity measurements [61] will allow minimum masses to be estimated for a larger number of rocky exoplanets with radius estimates already obtained from transits, thus allowing planet densities to be estimated.
 iv. ELTs and other ground-based platforms will greatly expand the list of rocky planets orbiting ultracool stars and characterize the atmospheres of some of them.
 v. WFIRST will demonstrate the coronagraph technology for a future direct imaging mission that would study Earth-like planets, if total mission cost can be limited.

**6. How to expand partnerships (interagency, international, public/private) to further the study of life's origin, evolution, distribution, and future in the universe**

 i. The search for life is now the 10th agency-wide objective of NASA. It includes the search both within and beyond the Solar System, and these have some overlapping objectives. The current organization of the Science Mission Directorate (Planetary, Astrophysics, Heliophysics, Earth Science) results in unintentional barriers to cross-disciplinary work required for the search for life. Given the emphasis on such activities at the agency level, SMD should consider restructuring or expanding its programs.
 ii. Increasing the rate at which innovative multi-disciplinary, inter-agency, and/or international projects are launched is a high priority. Physical centers/ hubs for short/ medium-term visitor programs (1-6 months) with daily talks and regular social events have a successful track record in jump-starting novel, high-impact new collaborations (Institutes for Advanced Study worldwide, Harvard's Radcliffe Institute, etc.).
 iii. New launch vehicles and optical design, fabrication, and testing technologies are needed to enable innovative, very large-aperture space-based telescopes or arrays. Much larger apertures will provide greater light-gathering power, essential for studying a larger sample of transiting exoplanets and increasing the sample of directly imaged planets. (Private enterprise builds most missions managed by NASA.)